\documentclass[final,5p,times,twocolumn]{elsarticle}

\usepackage[T1]{fontenc}
\usepackage[utf8]{inputenc}

\usepackage{amsmath}
\usepackage{amsfonts}
\usepackage{amssymb}
\usepackage{amsxtra}
\usepackage{array}
\usepackage{color}
\usepackage{dcolumn}
\usepackage{graphicx}
\usepackage{hepunits}
\usepackage{xspace}
\usepackage{CJKutf8}

\definecolor{purple}{rgb}{0.5,0,0.5}
\definecolor{blue}{rgb}{0.0,0,0.9}
\definecolor{prdblue}{rgb}{0.133,0.118,0.498}
\usepackage[colorlinks=true, pdfstartview=FitV, linkcolor=prdblue, citecolor= prdblue, urlcolor=prdblue]{hyperref}

\usepackage[mathscr,scaled=1.15]{urwchancal}
\DeclareFontFamily{OT1}{pzc}{}
\DeclareFontShape{OT1}{pzc}{m}{it}%
{<-> s * [1.15] pzcmi7t}{}
\DeclareMathAlphabet{\mathpzc}{OT1}{pzc}{m}{it}

\biboptions{sort&compress}

\journal{Fundamental Research}

\begin{document}
\begin{CJK*}{UTF8}{gbsn}
\begin{frontmatter}

\title{$\,$\\[-6ex]\hspace*{\fill}{\normalsize{\sf\emph{Preprint no}.\
NJU-INP 085/24}}\\[1ex]
Nucleon charge and magnetisation distributions: flavour separation and zeroes}

\author[NJU,INP,ECT]{Zhao-Qian Yao (姚照千)%
       $\,^{\href{https://orcid.org/0000-0002-9621-6994}{\textcolor[rgb]{0.00,1.00,0.00}{\sf ID}},}$}

\author[ECT]{Daniele Binosi%
    $\,^{\href{https://orcid.org/0000-0003-1742-4689}{\textcolor[rgb]{0.00,1.00,0.00}{\sf ID}},}$}

\author[NJU,INP]{Zhu-Fang~Cui (崔著钫)%
       $^{\href{https://orcid.org/0000-0003-3890-0242}{\textcolor[rgb]{0.00,1.00,0.00}{\sf ID}},}$}

\author[NJU,INP]{Craig D. Roberts%
       $^{\href{https://orcid.org/0000-0002-2937-1361}{\textcolor[rgb]{0.00,1.00,0.00}{\sf ID}},}$$^\ast$}

\address[NJU]{
School of Physics, \href{https://ror.org/01rxvg760}{Nanjing University}, Nanjing, Jiangsu 210093, China}
\address[INP]{
Institute for Nonperturbative Physics, \href{https://ror.org/01rxvg760}{Nanjing University}, Nanjing, Jiangsu 210093, China}
\address[ECT]{European Centre for Theoretical Studies in Nuclear Physics
            and Related Areas (\href{https://ror.org/01gzye136}{ECT*}), 
            Villa Tambosi, Strada delle Tabarelle 286, I-38123 Villazzano (TN), Italy\\[1ex]
%
\href{mailto:zyao@ectstar.eu}{zyao@ectstar.eu} (ZQY);
\href{mailto:binosi@ectstar.eu}{binosi@ectstar.eu} (DB);
\href{mailto:phycui@nju.edu.cn}{phycui@nju.edu.cn} (ZFC);
\href{mailto:cdroberts@nju.edu.cn}{cdroberts@nju.edu.cn} (CDR)
\\[1ex]
Date: 2024 November 06 \\[-6ex]
}

\cortext[1]{Corresponding author}
\fntext[1]{All authors contributed equally to this work.}

\begin{abstract}
A symmetry-preserving truncation of the quantum field equations describing hadron properties is used to deliver parameter-free predictions for all nucleon elastic electromagnetic form factors and their flavour separation to large values of momentum transfer, $Q^2$.  The proton electric form factor, $G_E^p$, possesses a zero, whereas that of the neutron, $G_E^n$, does not.  The difference owes to the behaviour of the Pauli form factor of the proton's singly-represented valence $d$-quark.  Consequently, $G_E^n>G_E^p$ on a material large-$Q^2$ domain.  These predictions can be tested in modern experiments.
\end{abstract}


\end{frontmatter}
\end{CJK*}


\section{Introduction}
\begin{figure}[b!]
\centerline{%
\includegraphics[clip, width=0.24\textwidth]{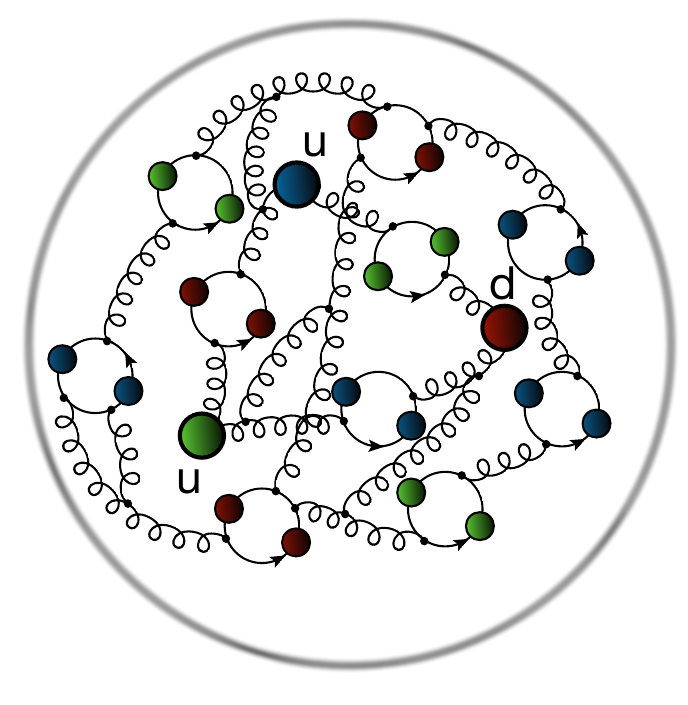}}
\caption{\label{Fproton}
Proton: two valence up ($u$) quark partons, one valence down ($d$) quark parton, and infinitely many gluon and sea-quark partons, drawn here as ``springs'' and closed loops, respectively.  The neutron is the proton's isospin partner, two $d$ quark partons, one $u$ quark parton, and glue and sea.}
\end{figure}

The proton is Nature's most fundamental bound state.  It is supposed to be explained by quantum chromodynamics (QCD), the Poincar\'e-invariant quantum non-Abelian gauge field theory that describes strong interactions in the Standard Model.  The QCD Lagrangian density is expressed in terms of gluon and quark partons (and ghosts, too, in many gauges) \cite{Marciano:1977su}.  In these terms, the proton consists of three valence-quark partons ($u+u+d$) and infinitely many gluon and sea-quark partons -- see Fig.\,\ref{Fproton}.  If science is to claim an understanding of Nature, then it must deliver a sound description of proton properties from QCD; not just its mass, but also its entire array of structural properties \cite{BESIII:2020nmeX, Anderle:2021wcyX, AbdulKhalek:2021gbhX, Quintans:2022utc, Carman:2023zke, Mokeev:2023zhq}.

%
The proton bound-state problem can be addressed in any approach that provides access to the three-quark six-point Schwinger function \cite{SW80, GJ81}.
Lattice-regularised QCD (lQCD) provides one such framework.
Modern applications are sketched in Ref.\,\cite[Sec.\,10]{FlavourLatticeAveragingGroupFLAG:2021npnX}.   Continuum Schwinger function methods (CSMs) provide another widely used approach to nucleon (proton, $p$, and neutron, $n$) structure \cite{Eichmann:2016yit, Burkert:2019bhp, Binosi:2022djx, Ding:2022ows, Ferreira:2023fva}.
Many such studies use a quark + dynamical diquark picture of the nucleon because it vastly simplifies the problem \cite{Barabanov:2020jvnX}.
Notwithstanding that, the approximations implicit in the simplification need checking and tighter links must be forged with QCD.
These things can be accomplished by beginning with an explicitly symmetry-preserving truncation of all quantum field equations (Dyson-Schwinger equations -- DSEs) relating to the nucleon bound-state problem.  The first study of this type was reported in Ref.\,\cite{Eichmann:2009qa}.

A highlight of proton structure studies this century is the collection of data that hints at the existence of a zero in the proton elastic electric form factor
\cite{Jones:1999rzX, Gayou:2001qdX, Punjabi:2005wqX, Puckett:2010acX, Puckett:2017fljX}.
(A zero in the transverse helicity amplitude associated with the proton $\to$ Roper transition has unambiguously been located \cite{Burkert:2019bhp}.)
This is complemented by the discovery of marked differences in the charge and magnetisation distributions of different valence-quark flavours ($u$ vs.\ $d$) within the proton \cite{Cates:2011pz, Wojtsekhowski:2020tlo}.
These features have attracted much attention \cite{Perdrisat:2006hj, Miller:2010nz, Holt:2012gg, Punjabi:2015bba, Brodsky:2020vcoX, Barabanov:2020jvnX}.
Modern and foreseen facilities will both obtain data that can check existing measurements and push empirical knowledge of all nucleon form factors to momentum transfers $Q^2 > 10\,$GeV$^2$ \cite{Gilfoyle:2018xsa, Wojtsekhowski:2020tlo}.  This prospect challenges theory to deliver predictions for all such form factors that extend far onto this domain in frameworks with a traceable connection to QCD.

Herein, we approach this challenge by using the rainbow-ladder (RL) truncation of all DSEs needed to calculate the 
matrix element from which nucleon elastic electromagnetic form factors can be extracted.
This is the leading-order in a symmetry-preserving, systematically-improvable scheme \cite{Binosi:2016rxz}.
Existing algorithms have limited the reach of such form factor calculations to $Q^2 \lesssim 4\,$GeV$^2$.
We extend the results to $Q^2 \gtrsim 12\,$GeV$^2$ using the statistical Schlessinger point method (SPM)
\cite{Cui:2022fyr, Binosi:2022ydc, Cui:2022dcm}, which may also be called a statistical multi-point Pad\'e approximant scheme.  The SPM is grounded in analytic function theory.
It is free from practitioner-induced bias; hence, delivers objective analytic continuations with quantitatively reliable uncertainties.


\section{Methods and Tools}
\subsection{Nucleon  Bound State}
\label{NBS}
\begin{figure}[t]
\centerline{%
\includegraphics[clip, width=0.44\textwidth]{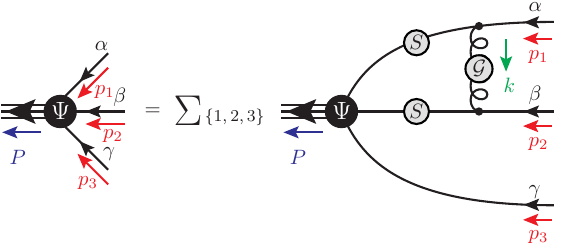}}
\caption{\label{FigFaddeev}
Faddeev equation. 
Filled circle: Faddeev amplitude, $\Psi$, the matrix-valued solution, which involves 128 independent scalar functions.
Spring: dressed-gluon interaction that mediates quark+quark scattering, Eqs.\,\eqref{EqRLInteraction}, \eqref{defcalG}.
Solid line: dressed-quark propagator, $S$, calculated from the rainbow gap equation.
Lines not adorned with a shaded circle are amputated.  Isospin symmetry is assumed.}
\end{figure}

The RL truncation nucleon Faddeev equation is drawn in Fig.\,\ref{FigFaddeev}.  Discussions of the formulation and solution of this linear, homogeneous integral equation are provided, \emph{e.g}., in Ref.\, \cite{Wang:2018kto, Qin:2018dqp}.  The key element is the quark + quark scattering kernel, for which the RL truncation is obtained by writing \cite{Maris:1997tm}:
{\allowdisplaybreaks
%
\begin{equation}
\label{EqRLInteraction}
\mathscr{K}_{tu}^{rs}(k)  =
\tilde{\mathpzc G}(y)
 [i\gamma_\mu\frac{\lambda^{a}}{2} ]_{ts} [i\gamma_\nu\frac{\lambda^{a}}{2} ]_{ur}
 T_{\mu\nu}(k)\,,
\end{equation}
$k^2 T_{\mu\nu}(k) = k^2 \delta_{\mu\nu} - k_\mu k_\nu$,  $y=k^2$.  The tensor structure specifies Landau gauge, used because it is a fixed point of the renormalisation group and that gauge for which corrections to RL truncation are least significant \cite{Bashir:2009fv}.
In Eq.\,\eqref{EqRLInteraction}, $r,s,t,u$ represent colour, spinor, and flavour matrix indices (as necessary).
}

\begin{figure*}[t]
\centerline{%
\includegraphics[clip, width=0.86\textwidth]{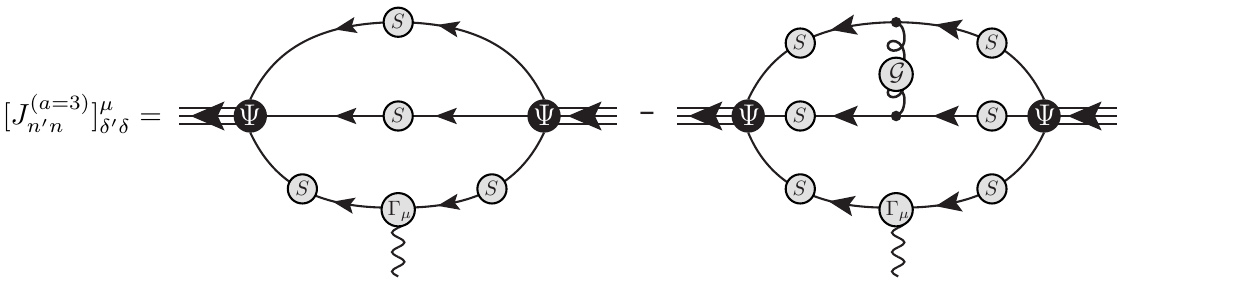}}
\caption{\label{FigCurrent}
Since the nucleon has three valence quarks, the complete nucleon electromagnetic current has three terms: $J_\mu(Q) = \sum_{a=1,2,3} J_\mu^a(Q)$; but using symmetries, one can readily obtain the $a=1,2$ components once the $a=3$ component is known \cite[Appendix~B]{Eichmann:2011pv}.
%
$\delta$, $\delta^\prime$ are spinor indices and $n$, $n^\prime$ are isospin indices.
$\Gamma_\mu$ is the dressed-photon+quark vertex, which can be obtained, \emph{e.g}., following Ref.\,\cite{Xu:2019ilh}.
%
}
\end{figure*}

A realistic form of ${\mathpzc G}_{\mu\nu}(y)$ is explained in Refs.\,\cite{Qin:2011dd, Binosi:2014aea}:
\begin{align}
\label{defcalG}
 \tilde{\mathpzc G}(y) & =
 \frac{8\pi^2}{\omega^4} D e^{-y/\omega^2} + \frac{8\pi^2 \gamma_m \mathcal{F}(y)}{\ln\big[ \tau+(1+y/\Lambda_{\rm QCD}^2)^2 \big]}\,,
\end{align}
where $\gamma_m=12/25$, $\Lambda_{\rm QCD} = 0.234\,$GeV, $\tau={\rm e}^2-1$, and ${\cal F}(y) = \{1 - \exp(-y/\Lambda_{\mathpzc I}^2)\}/y$, $\Lambda_{\mathpzc I}=1\,$GeV.
%
%
We employ a mass-independent (chiral-limit) momentum-subtraction renormalisation scheme \cite{Chang:2008ec}. 

Widespread use has shown \cite{Ding:2022ows} that interactions in the class containing Eqs.\,\eqref{EqRLInteraction}, \eqref{defcalG} can serve to unify the properties of many systems.
Contemporary studies employ $\omega = 0.8\,$GeV \cite{Xu:2022kng}.  Then, with $\omega D = 0.8\,{\rm GeV}^3$ and renormalisation point invariant quark current mass $\hat m_u = \hat m_d = 6.04\,$MeV, which corresponds to a one-loop mass at $\zeta=2\,$GeV of $4.19\,$MeV, the following predictions are obtained: pion mass $m_\pi = 0.14\,$GeV; nucleon mass $m_N=0.94\,$GeV; and pion leptonic decay constant $f_\pi=0.094\,$GeV.  These values align with experiment \cite{ParticleDataGroup:2024cfkX}.
%
%
When the product $\omega D$ is kept fixed, physical observables remain practically unchanged under $\omega \to (1\pm 0.2)\omega$ \cite{Qin:2020rad}.

All subsequent calculations are parameter-free.  The interaction involves one parameter and there is a single quark current-mass.  Both quantities are now fixed.

Before continuing, it is worth providing additional context for the interaction in Eq.\,\eqref{defcalG} by noting that, following Ref.\,\cite{Qin:2011dd}, one may draw a connection between $\tilde{\mathpzc G}$
and QCD's process-independent effective charge, discussed in Refs.\,\cite{Cui:2019dwv, Brodsky:2024zev}.
That effective charge is characterised by an infrared coupling value $\hat\alpha(0)/\pi = 0.97(4)$ and a gluon mass-scale $\hat m_0 = 0.43(1)\,$GeV determined in a combined continuum and lattice analysis of QCD's gauge sector \cite{Cui:2019dwv}.
The following values are those of analogous quantities inferred from Eq.\,(2): 
\begin{equation}
\label{DiscussInteraction}
\alpha_{\mathpzc G}(0)/\pi = 1.45\,,\quad m_{\mathpzc G} = 0.54\,{\rm GeV}\,.
\end{equation}
They agree tolerably with the QCD values, especially if one recalls that earlier, less well informed versions of the RL interaction yielded $\alpha_{\mathpzc G}(0)/\pi \approx 15$, \emph{i.e}., a value ten-times larger~\cite{Qin:2011dd}.

\subsection{Nucleon electromagnetic current}
\label{NEM}
Working with the solution of the Faddeev equation in Fig.\,\ref{FigFaddeev}, the interaction current drawn in Fig.\,\ref{FigCurrent} is necessary and sufficient to deliver a photon + nucleon interaction that is consistent with all relevant Ward-Green-Takahashi identities; hence, \emph{inter alia}, ensures electromagnetic current conservation \cite[Sec.\,III.A]{Eichmann:2011vu}.  The current can be written as follows ($N=p,n$):
\begin{align}
J_\mu^N(Q) & = ie \Lambda_+(p_f)
[ F_1^N(Q^2) \gamma_\mu \nonumber \\
& \quad + \frac{1}{2 m_N} \sigma_{\mu\nu} Q_\nu F_2^N(Q^2) ]
\Lambda_+(p_i)  \label{NucleonCurrent}
\end{align}
where $e$ is the positron charge,
the incoming and outgoing nucleon momenta are $p_{i,f}$, $Q=p_f-p_i$,
$\Lambda_+(p_{i,f})$ are positive-energy nucleon-spinor projection operators,
and $F_{1,2}^N$ are the Dirac and Pauli form factors.

The nucleon charge and magnetisation distributions are ($\tau = Q^2/[4 m_N^2]$) \cite{Sachs:1962zzc}:
\begin{equation}
G_E^N  = F_1^N - \tau F_2^N\,,
\quad G_M^N  = F_1^N + F_2^N \,.
\label{Sachs}
\end{equation}
Magnetic moments and radii are obtained therefrom using standard definitions:
$\mu_N = G_M^N(Q^2=0)$\,;
\begin{align}
\label{StandardDef}
\langle r_F^2\rangle^N & = \left. -6 \frac{d \ln G_F^N(Q^2)}{dQ^2}\right|_{Q^2=0}\,,
\end{align}
$F=E$, $M$, except $\langle r_E^2\rangle^n = -6 G_E^{n\prime}(Q^2)|_{Q^2=0}$ because $G_E^{n}(0)=0$.

Numerical methods for solving sets of coupled gap, Bethe-Salpeter, and Faddeev equations are described, \emph{e.g}., in Refs.\,\cite{Maris:1997tm, Maris:2005tt, Krassnigg:2009gd, Qin:2018dqp}.  Exploiting these schemes, we solved all equations relevant to calculation of the current in Eq.\,\eqref{NucleonCurrent} and computed the current itself, thereby arriving at predictions for the form factors in Eq.\,\eqref{Sachs}.

%

A technical remark is appropriate here.
The Faddeev equation solution depends on two relative momenta, $p$, $q$, and the nucleon total momentum, $P$.
This leads to a dependence on three angular variables defined via the inner products $p\cdot q$, $p\cdot P$, $q\cdot P$.
In solving the equation, eight Chebyshev polynomials are used to express the dependence on each angle \cite{Maris:1997tm}.
This enables evaluation of $\Psi$ at any required integration point in either the Faddeev equation or the current.
$P$ is a complex-valued (timelike) vector, $P^2=-m_N^2$, whereas $Q$ is spacelike.
So, when evaluating the current, the integrand sample points are typically in the complex plane.
This leads to oscillations whose amplitudes grow with $Q^2$.
Increasing the number of Chebyshev polynomials and quadrature points is effective on $Q^2\lesssim 4\,$GeV$^2$.
At larger $Q^2$ values, however, this brute force approach fails to deliver accurate results.

In order to obtain predictions on $Q^2\gtrsim 4\,$GeV$^2$, we extrapolate using the SPM, whose properties and accuracy are explained and illustrated elsewhere -- see Refs.\,\cite{Cui:2022fyr, Binosi:2022ydc, Cui:2022dcm} and citations therein and thereof.
The SPM is based on the Pad\'e approximant.
It accurately reconstructs a function in the complex plane within a radius of convergence determined by that one of the function's branch points which lies closest to the real domain that provides the sample points.  Modern implementations introduce a statistical element; so, the extrapolations come with an objective and reliable estimate of uncertainty.

As noted above, numerous demonstrations of the SPM's accuracy and reliability are available.
Herein we highlight (\emph{i}) that provided in connection with the proton radius puzzle \cite[Supplemental Material]{Cui:2021vgm}, which shows that the SPM accurately reproduces known radii from nine distinct representations of low-energy electron + proton scattering data and (\emph{ii}) an application to the search for exotic hadrons, in which the SPM was shown to reliably reproduce results obtained using five distinct models that were employed to perform a combined analysis of different partial waves in the decays $J/\psi \to \gamma \pi^0 \pi^0$ and $J/\psi \to \gamma K_S^0 K_S^0$ \cite[Sec.\,3]{Binosi:2022ydc}.
It is also worth highlighting that in these diverse applications, the SPM was applied without modification.  It is a truly robust tool.

Our SPM extrapolations of the form factors onto $Q^2 \gtrsim 4\,$GeV$^2$ are developed as follows.
\begin{description}
\item[Step 1]
For each function considered, we produce $N=40$ directly calculated function values spaced evenly on $Q^2\lesssim 4\,$GeV$^2$.
\item[Step 2]
From that set, $M_0=14$ points are chosen at random,
the usual SPM continued fraction interpolation is constructed,
and that function is extrapolated onto $Q^2/{\rm GeV}^2 \in [4,12]$.
The curve is retained so long as it is singularity free.
\item[Step 3]
This is repeated with another set of $M_0$ randomly chosen points.
Steps 2 and 3 admit $\approx 5\times 10^{10}$ independent extrapolations.
\item[Step 4]
One continues with 2 and 3 until $n_{M_0}=500\,$ smooth extrapolations are obtained.
\item[Step 5]
Steps 2 and 3 are repeated for $M=\{M_0+ 2 i | i=1,\ldots,6\}$
\item [Step 6]
At this point, one has $3\,000$ statistically independent extrapolations.
\end{description}
Working with these extrapolations, then at each value of $Q^2$, we record the mean value of all curves as the central prediction and report as the uncertainty the function range which contains 68\% of all the extrapolations -- this is a $1\sigma$ band.

\begin{table}[t]
\caption{\label{TabcfExpt}
Static properties: magnetic moments in nuclear magnetons and radii-squared in fm$^2$, calculated using 
conventional definitions -- Sec.\,\ref{NEM}.
Empirical values from Ref.\,\cite[PDG]{ParticleDataGroup:2024cfkX}.
The column ``SPM'' lists radii extracted from experimental data using the SPM
\cite{Cui:2022fyr}. 
 }
\begin{center}
\begin{tabular*}
{\hsize}
{
l@{\extracolsep{0ptplus1fil}}
|l@{\extracolsep{0ptplus1fil}}
l@{\extracolsep{0ptplus1fil}}
l@{\extracolsep{0ptplus1fil}}}\hline\hline
 & herein & Exp. & SPM  \\\hline
 $\mu_p\ $ & $\phantom{-}2.23\ $ &$\phantom{-}2.793\ $  & \\
 $\mu_n\ $ & $-1.33\ $ & $-1.913\ $ & \\
 $\langle r_E^2 \rangle^p$ & $\phantom{-}0.788\ $ & $\phantom{-}0.7070(7)\ $ & $0.717(14)\ $ \\
 $\langle r_E^2 \rangle^n$ & $-0.0621\ $ & $-0.1160(22)\ $ &  \\
 $\langle r_M^2 \rangle^p$ & $\phantom{-}0.672\ $ & $\phantom{-}0.72(4)\ $ & $0.667(44)\ $ \\
 $\langle r_M^2 \rangle^n$ & $\phantom{-}0.661\ $ & $\phantom{-}0.75(2)\ $ &
 \\\hline\hline
\end{tabular*}
\end{center}
\end{table}

\section{Results and Discussion}
\subsection{Nucleon form factors}
\label{NFF}
\begin{figure}[t]
\vspace*{1.2em}

\leftline{\hspace*{0.5em}{\large{\textsf{A}}}}
\vspace*{-5ex}
\includegraphics[width=0.41\textwidth]{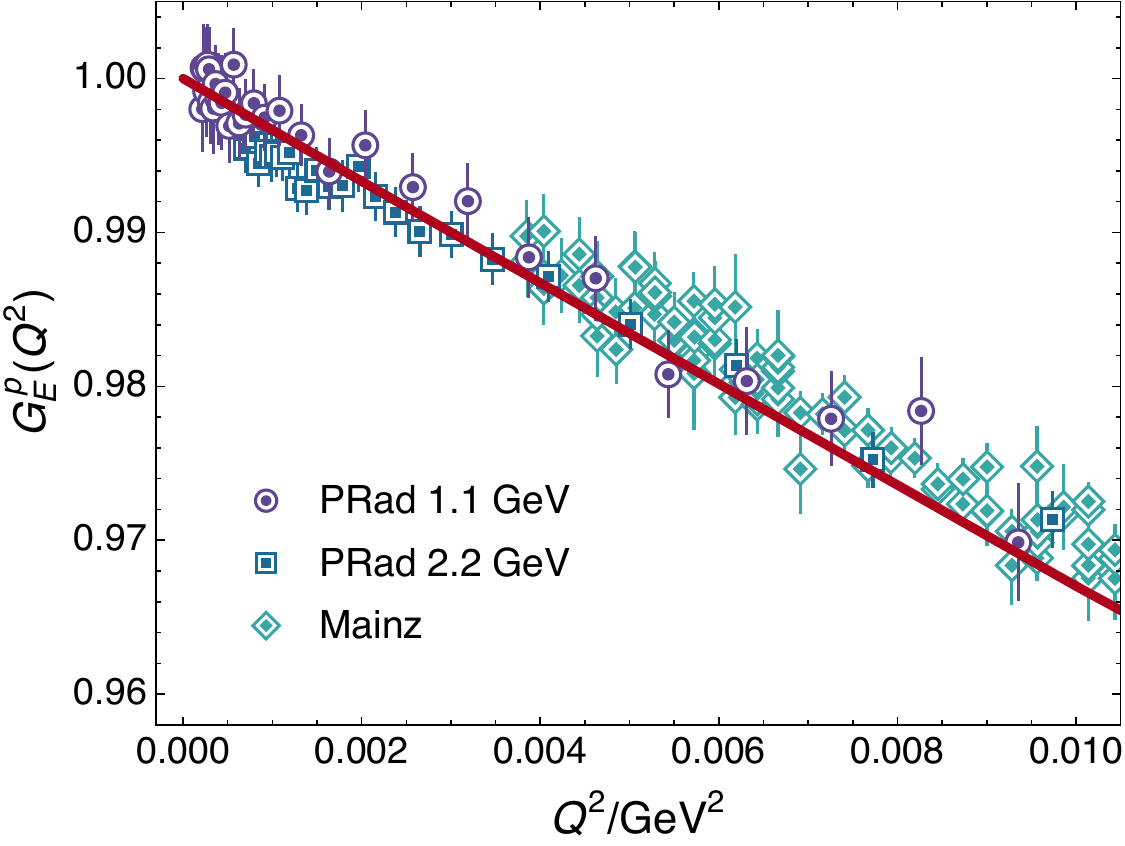}
\vspace*{0.1ex}
\leftline{\hspace*{0.5em}{\large{\textsf{B}}}}
\vspace*{-5ex}
\includegraphics[width=0.41\textwidth]{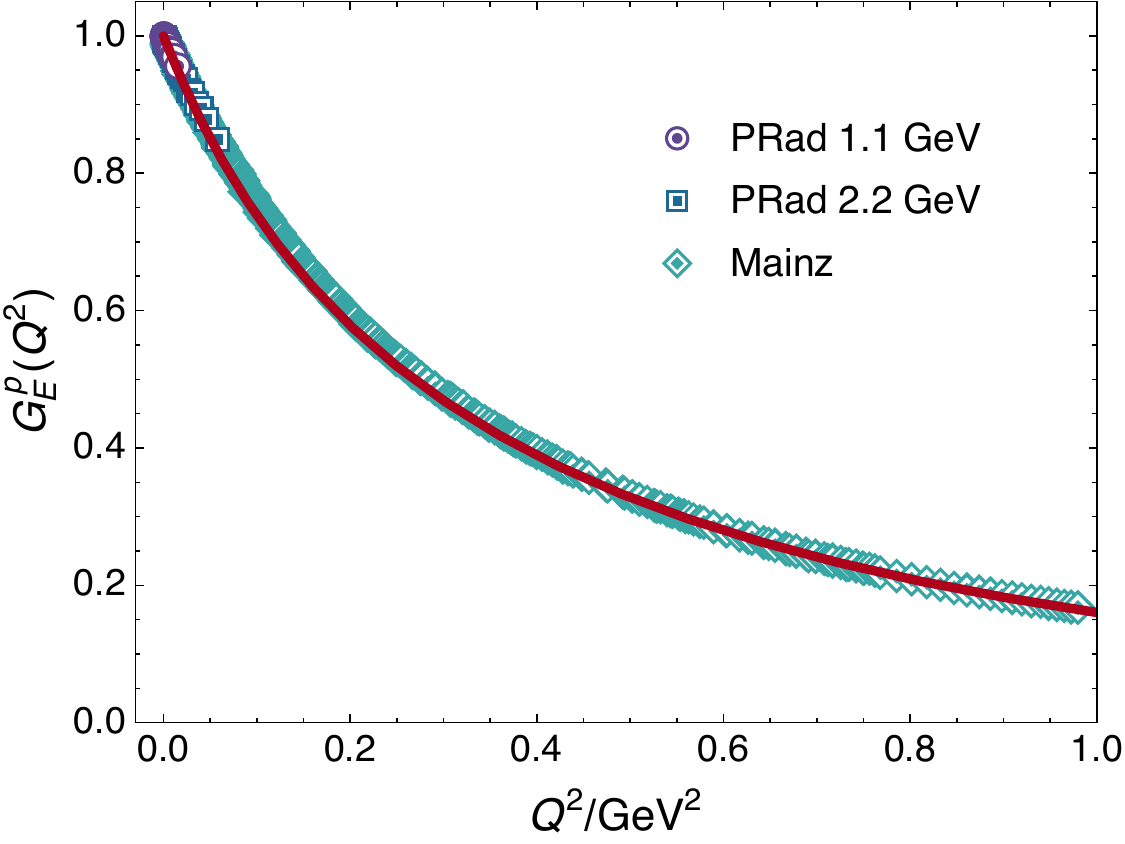}
\vspace*{3.5ex}

\caption{\label{lowQ2}
Low $Q^2$ behaviour of proton electric form factor: solid red line -- result obtained herein;
data from Refs.\,\cite[Mainz]{Bernauer:2010wmX} and \cite[PRad]{Xiong:2019umfX}.}
\end{figure}

Predictions for nucleon static (low $Q^2$) properties are collected in Table~\ref{TabcfExpt}.  In size, the magnetic moments are $\sim 25$\% too small.  This is a failing of RL truncation, which produces a photon+quark vertex whose dressed-quark anomalous magnetic moment term is too weak.  It is corrected in higher-order truncations \cite{Chang:2010hb}.  Such corrections have been implemented in studies of mesons \cite{Xu:2022kng}.  It may be possible to adapt this approach to baryons.  Concerning the other entries in Table~\ref{TabcfExpt}, the agreement with experiment is reasonable.
In particular,
our analysis delivers fair agreement with extant low-$Q^2$ precision data on electron + proton scattering -- see Fig.\,\ref{lowQ2}; and the prediction $\langle r_E^2 \rangle^p > \langle r_M^2 \rangle^p$ accords with SPM analyses of existing form factor measurements \cite{Cui:2022fyr}. 

\begin{figure}[t]
\vspace*{1.2em}

\leftline{\hspace*{0.5em}{\large{\textsf{A}}}}
\vspace*{-5ex}
\includegraphics[width=0.40\textwidth]{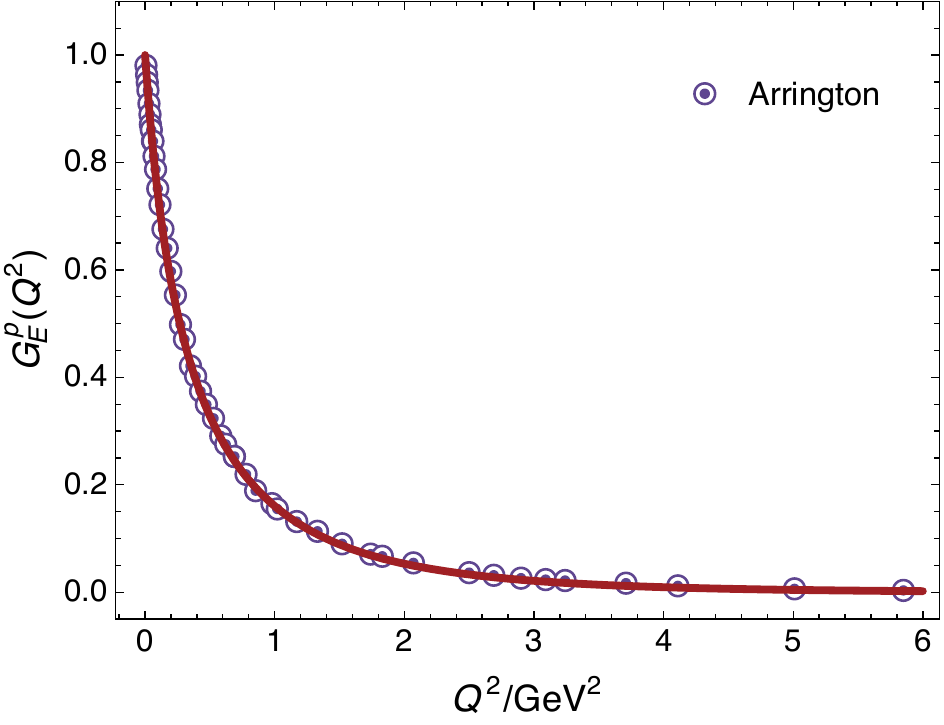}
\vspace*{0.1ex}
\leftline{\hspace*{0.5em}{\large{\textsf{B}}}}
\vspace*{-5ex}
\includegraphics[width=0.40\textwidth]{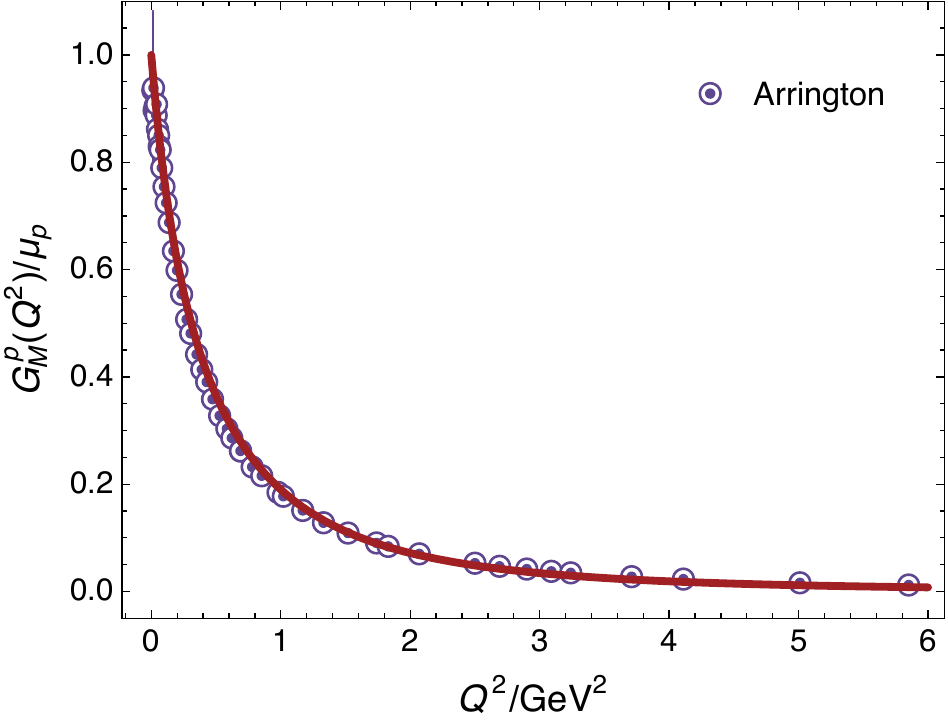}
\vspace*{3.5ex}

\caption{\label{figGEMnpA}
Proton electromagnetic form factors: solid red line -- results obtained herein.
Experimental data taken from compilation in Ref.\,\cite{Arrington:2007ux}.}
\end{figure}

\begin{figure}[t]
\vspace*{1.2em}

\leftline{\hspace*{0.5em}{\large{\textsf{A}}}}
\vspace*{-5ex}
\includegraphics[width=0.41\textwidth]{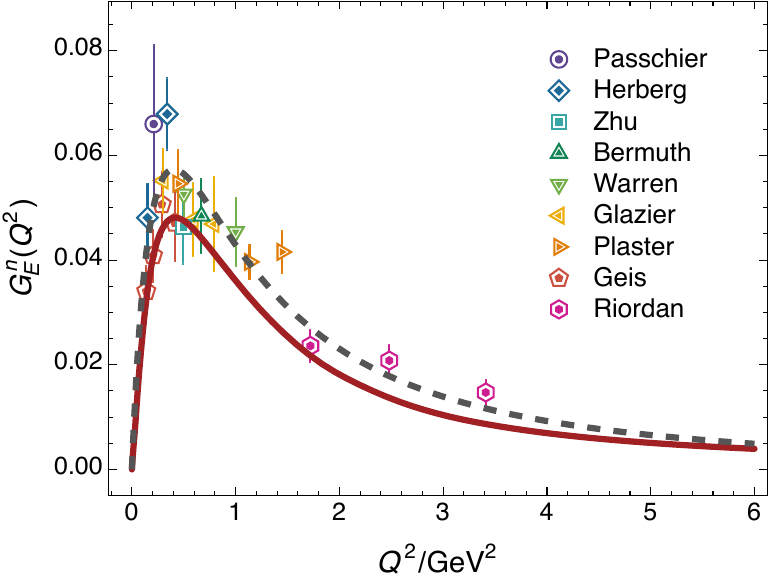}
\vspace*{0.1ex}
\leftline{\hspace*{0.5em}{\large{\textsf{B}}}}
\vspace*{-5ex}
\includegraphics[width=0.41\textwidth]{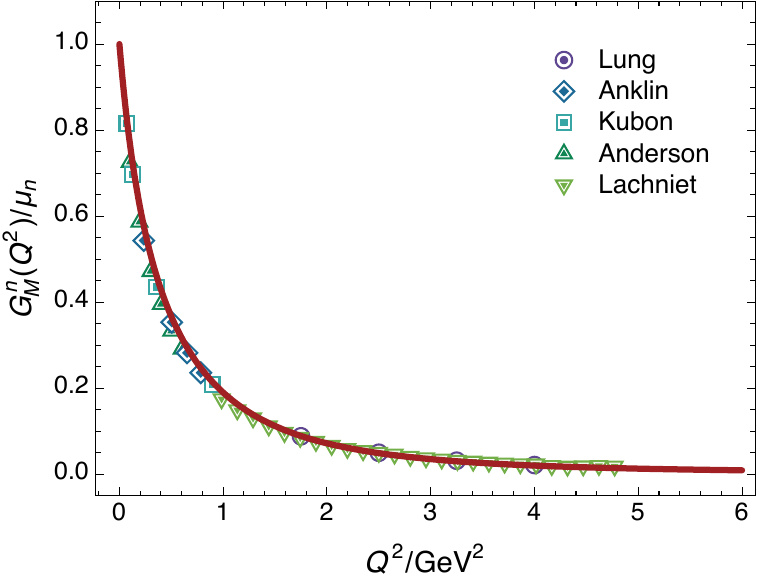}
\vspace*{3.5ex}
%
\caption{\label{figGEMnpB}
Neutron electromagnetic form factors.
Solid red curve -- results obtained herein.
Dashed black curve in panel~A -- Ref.\,\cite[Kelly]{Kelly:2004hm} parametrisation of data.
$G^n_E$ experimental data: Refs.\,\cite{Passchier:1999cj, Herberg:1999ud, E93026:2001css, Bermuth:2003qh, Warren:2003ma, Glazier:2004ny, Plaster:2005cx, BLAST:2008bub, Riordan:2010id}.
$G^n_M$ data: Refs.\,\cite{Lung:1992buX, Anklin:1998aeX, Kubon:2001rj, JeffersonLabE95-001:2006dax, CLAS:2008idi}}
%
\end{figure}

As displayed in Figs.\,\ref{figGEMnpA}, \ref{figGEMnpB}, the Faddeev equation prediction for the overall $Q^2$ dependence of each nucleon 
form factor agrees well with data
\cite{Arrington:2007ux, Passchier:1999cj, Herberg:1999ud, E93026:2001css, Bermuth:2003qh, Warren:2003ma, Glazier:2004ny, Plaster:2005cx, BLAST:2008bub, Riordan:2010id, Lung:1992buX, Anklin:1998aeX, Kubon:2001rj, JeffersonLabE95-001:2006dax, CLAS:2008idi}.
Even $G_E^n(Q^2)$ is a fair match, despite its sensitivity to details of the neutron wave function, especially as expressed in $F_1^n$ -- see, \emph{e.g}., Refs.\,\cite{Eichmann:2016yit, Segovia:2014aza}.

{\allowdisplaybreaks
It is worth quantifying the above remarks by comparing the predictions in Figs.\,\ref{figGEMnpA}, \ref{figGEMnpB} with the parametrisations of data provided in Ref.\,\cite[Kelly]{Kelly:2004hm}.  A useful measure is the relative ${\mathpzc L}_1$ difference: $\Delta_F^N= 2 [\delta_-]_F^N/[\delta_+]_F^N$, where
\begin{align}
[\delta_\mp ]_F^N & = \int_0^{10\,{\rm GeV}^2} \!\! dQ^2 \nonumber \\
& \qquad | {\rm Prediction}_F^N(Q^2) \mp  {\rm Kelly}_F^N(Q^2)|\,.
\end{align}
The upper bound is effectively that employed in Ref.\,\cite{Kelly:2004hm}.
The results are:
\begin{equation}
\label{DataAgreement}
\begin{array}{l|c|c|c|c}
 & G_E^p & G_M^p & G_E^n & G_M^n\\\hline
\Delta_F^N(\%) & 4.9 & 7.2 & 21 & 4.0
\end{array}
\end{equation}}

Evidently, the parameter-free Faddeev equation predictions are practically indistinguishable from the data fits \cite{Kelly:2004hm}, except in the case of $G_E^n$, which, in the mean, lies systematically below the fit by $\approx 20$\%.
These features are also illustrated in Figs.\,\ref{figGEMnpA}, \ref{figGEMnpB}.
Regarding $G_E^p$, $G_M^p/\mu_p$, $G_M^n/\mu_n$, within line width, the data parametrisations are indistinguishable from our predictions -- so, not drawn.
The parametrisation is drawn in Fig.\,\ref{figGEMnpB}A, making manifest the $\approx 20$\% underestimate of $G_E^n$.

\subsection{Form Factor Ratios}
\label{NFFR}
\begin{figure}[t]
\vspace*{1.2em}

\leftline{\hspace*{0.5em}{\large{\textsf{A}}}}
\vspace*{-5ex}
\includegraphics[clip, width=0.41\textwidth]{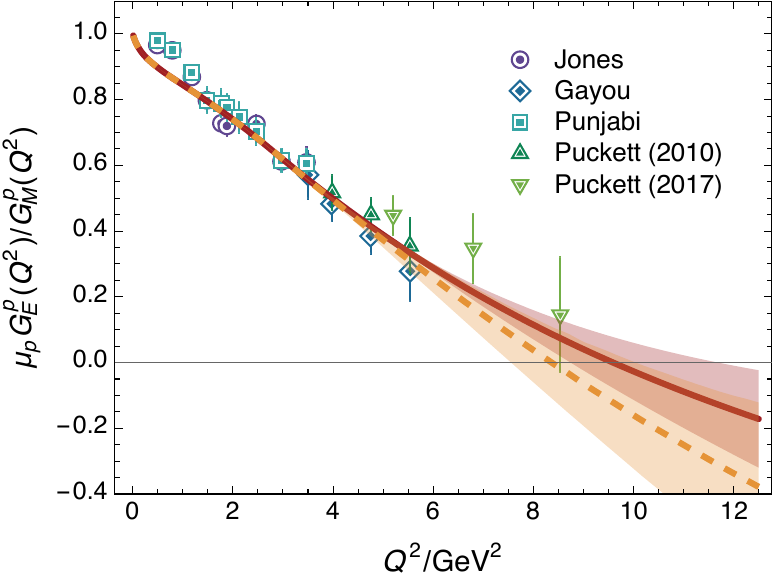}
\vspace*{1ex}
\leftline{\hspace*{0.5em}{\large{\textsf{B}}}}
\vspace*{-5ex}
\includegraphics[clip, width=0.41\textwidth]{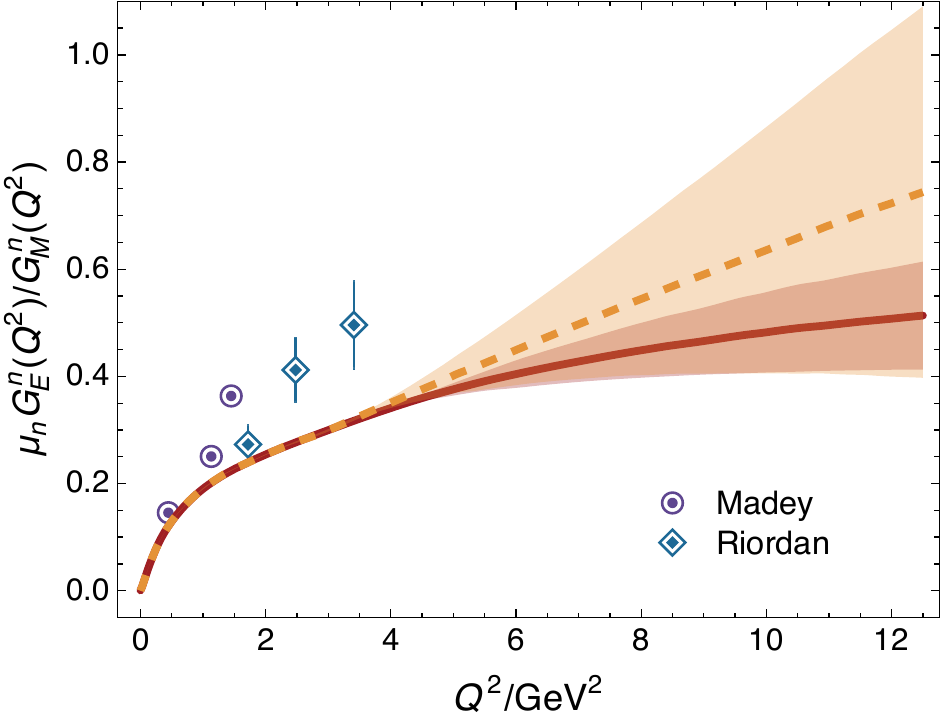}
\vspace*{4.5ex}
\caption{\label{mupGEpGMp}
%
{\sf Panel A}: $\mu_p G_E^p/G_M^p$.
{\sf Panel B}: $\mu_n G_E^n/G_M^n$.
SPM I -- dashed orange curve within like-coloured band; and SPM II -- solid red curve within like-coloured band.
Data: proton -- Refs.\,\cite{Jones:1999rzX, Gayou:2001qdX, Punjabi:2005wqX, Puckett:2010acX, Puckett:2017fljX}; and neutron -- Refs.\,\cite{Madey:2003av, Riordan:2010id}.
%
}
\end{figure}

{\allowdisplaybreaks
It is appropriate here to stress that $G_M^{p,n}(Q^2)/\mu_{p,n}$ agree well with experiment.  This is important in connection with the prediction for $\mu_p G_E^p(Q^2)/G_M^p(Q^2)$ drawn in Fig.\,\ref{mupGEpGMp}A.
Directly calculated Faddeev equation results are available on $Q^2\lesssim 4\,$GeV$^2$.
Thereafter, we calculate two sets of SPM results:
(\emph{I}) ratio formed from curves obtained via independent SPM analyses of $G_{E,M}^p$;
(\emph{II}) SPM analysis of the ratio $\mu_p G_E^p/G_M^p$ obtained on the directly accessible domain.
Both methods yield compatible results and
agree with all available data within mutual uncertainties.
Significantly, a zero is predicted in $G_E^p$:
\begin{subequations}
\label{ZeroLocate}
\begin{align}
\mbox{SPM~I:} & \quad Q_{G_E^p{\rm -zero}}^2 = 8.37 ^{+1.68}_{-0.81}\,{\rm GeV}^2\,, \\
\mbox{SPM~II:} & \quad Q_{G_E^p{\rm -zero}}^2 = 9.59 ^{+2.09}_{-0.85}\,{\rm GeV}^2\,.
\end{align}
\end{subequations}
Being compatible, they can be averaged, with the result:
\begin{equation}
\label{GEpzero}
Q_{G_E^p{\rm -zero}}^2 = 8.86^{+1.93}_{-0.86}\,{\rm GeV}^2.
\end{equation}
}

Notably, we have verified the suggestion made elsewhere \cite{Cloet:2013gva} that if the quark + quark interaction is modified such that dressed quarks more rapidly become parton like, then $Q_{G_E^p{\rm -zero}}^2$ is shifted to a larger value.  The location of the zero in $G_E^p$ is thus confirmed to be a sensitive expression of gauge sector dynamics and emergent hadron mass \cite{Binosi:2022djx, Ding:2022ows, Ferreira:2023fva}.

We depict the Faddeev equation prediction for  $\mu_n G_E^n(Q^2)/G_M^n(Q^2)$ in Fig.\,\ref{mupGEpGMp}B.
The agreement with data is fair and the trend is correct.
Given that our prediction delivers a good description of $G_M^n(Q^2)/\mu_n$, the quantitative mismatch owes to the imperfect description of $G_E^n$ that was described above.
No signal is found for a zero in $\mu_n G_E^n(Q^2)/G_M^n(Q^2)$.
It follows that there is a $Q^2$ domain upon which the charge form factor of the neutral neutron is larger than that of the positively charged proton.  It begins at $Q^2 = 4.66 ^{+0.18}_{-0.13}$GeV$^2$.

At first glance, the absence of a zero in $\mu_n G_E^n(Q^2)/G_M^n(Q^2)$ conflicts with the other existing Poincar\'e-invariant study of nucleon form factors at large $Q^2$, which employs a quark+diquark approach \cite{Cui:2020rmu}.
However, that study locates the zero at $Q_{G_E^n{\rm -zero}}^2 = 20.1^{+10.6}_{-3.5}\,$GeV$^2$, \emph{i.e}., an uncertain location beyond the range of foreseeable measurements.

Naturally, given the simplicity of the quark + diquark approach, some differences should be expected between our predictions and the results in Ref.\,\cite{Cui:2020rmu}.  Comparisons are nevertheless worthwhile because they can inform the improvement of both approaches.  Indeed, given that its simplicity enables straightforward application to a wide range of problems, the value of a refined quark + diquark approach should not be underestimated.
It is important, therefore, to observe that the predictions herein and those in Ref.\,\cite{Cui:2020rmu} are largely in semiquantitative agreement, even though apparently minor differences are amplified at large $Q^2$.  This means that an efficacious refinement of the quark + diquark picture is achievable.


\begin{figure}[t]
\vspace*{1.2em}

\leftline{\hspace*{0.5em}{\large{\textsf{A}}}}
\vspace*{-5ex}
\includegraphics[clip, width=0.41\textwidth]{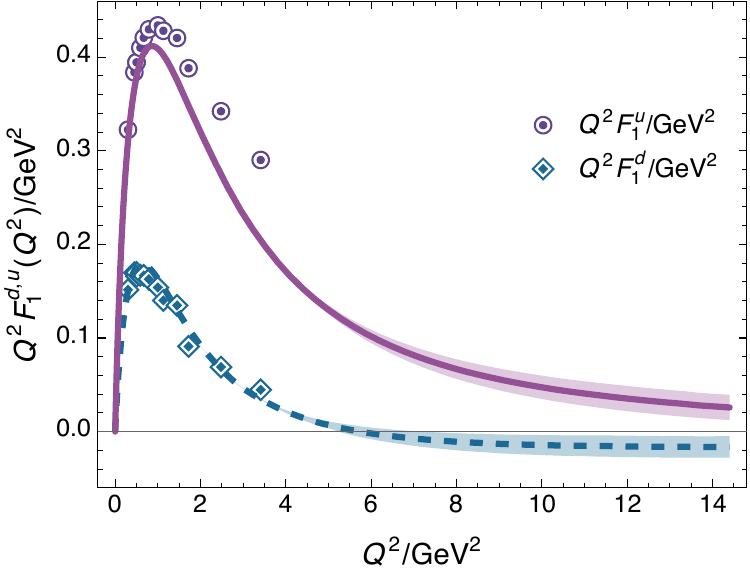}
\vspace*{1ex}
\leftline{\hspace*{0.5em}{\large{\textsf{B}}}}
\vspace*{-5ex}
\includegraphics[clip, width=0.41\textwidth]{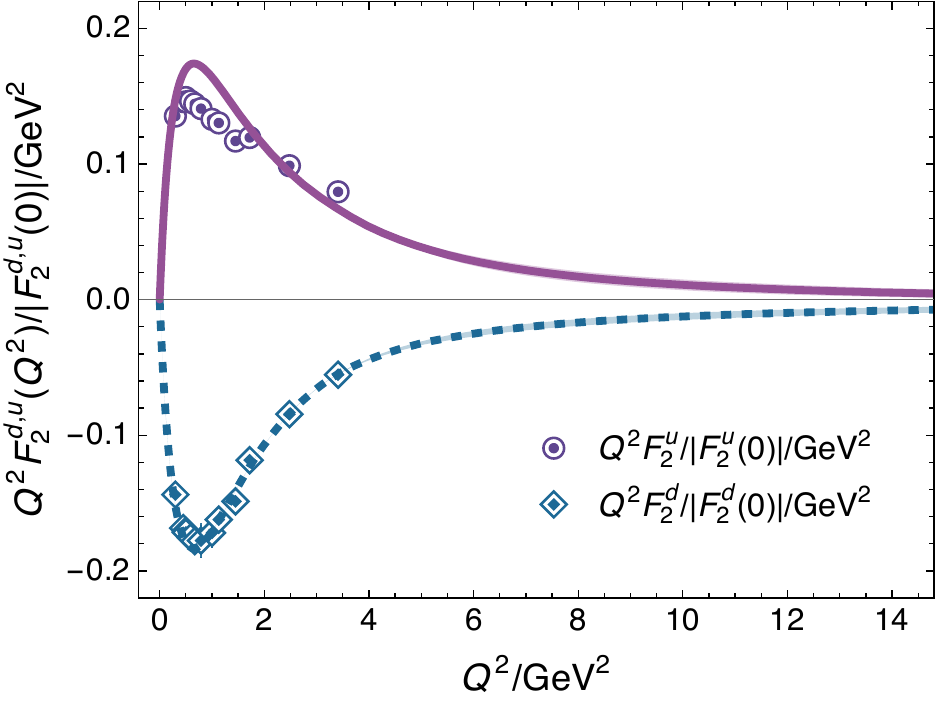}
\vspace*{4.5ex}
\caption{\label{FlavSep}
Flavour separation of proton form factors: $Q^2 F_1^{d,u}(Q^2)$ ({\sf Panel A}); and $Q^2 F_2^{d,u}(Q^2)/|F_2^{d,u}(0)/|$ ({\sf Panel B}).
Data: Ref.~\cite{Cates:2011pz}.
}
\end{figure}

\subsection{Flavour Separation}
\label{NFFFS}
Supposing one can neglect strange quark contributions to nucleon form factors, which is a good approximation \cite{Shanahan:2014tja}, then a flavour separation is possible using the following identities:
\begin{equation}
\label{FlavourSep}
F_{i}^u = 2 F_{i}^p + F_{i}^{n}, \;
F_{i}^d = F_{i}^{p} + 2 F_{i}^{n}, \; i=1,2\,.
\end{equation}
Current conservation and valence-quark number entail
$F_{1}^{u}(Q^2=0)=2$, $F_{1}^{d}(Q^2=0)=1$.

Our parameter-free predictions for these form factors are drawn in Fig.\,\ref{FlavSep}.
They deliver good agreement with available data.
\emph{N.B}. To account for the RL truncation underestimate of nucleon magnetic moments, the Pauli form factors in Fig.\,\ref{FlavSep}B -- both experiment and theory -- are normalised by the magnitude of their $Q^2=0$ values.
Regarding Fig.\,\ref{FlavSep}A, it is worth highlighting that the apparent mismatch between our prediction for $F_1^u$ and larger $Q^2$ data is somewhat misleading owing to amplification via $Q^2$ multiplication.  On the displayed domain, the true relative ${\mathpzc L}_1$ difference between prediction and data is just 6\%.
There is room for improvement in the RL treatment of the three-valence-body problem, but it does provide a reliable foundation.  Notably, a zero is projected in $F_1^d$ at
\begin{equation}
Q^2_{F_1^d{\rm -zero}} =5.73^{+1.46}_{-0.49}\,{\rm GeV}^2.
\end{equation}
This matches the result obtained in the quark+diquark picture \cite{Cui:2020rmu}: $Q^2=7.0^{+1.1}_{-0.4}\,$GeV$^2$.

No signal is found for a zero in any other form factor in Fig.\,\ref{FlavSep}.
The quark + diquark picture produces an uncertain zero in $F_2^d$ at very large momentum transfer: $Q^2=12.0^{+3.9}_{-1.7}\,$GeV$^2$.

\begin{figure}[t]
\includegraphics[width=0.41\textwidth]{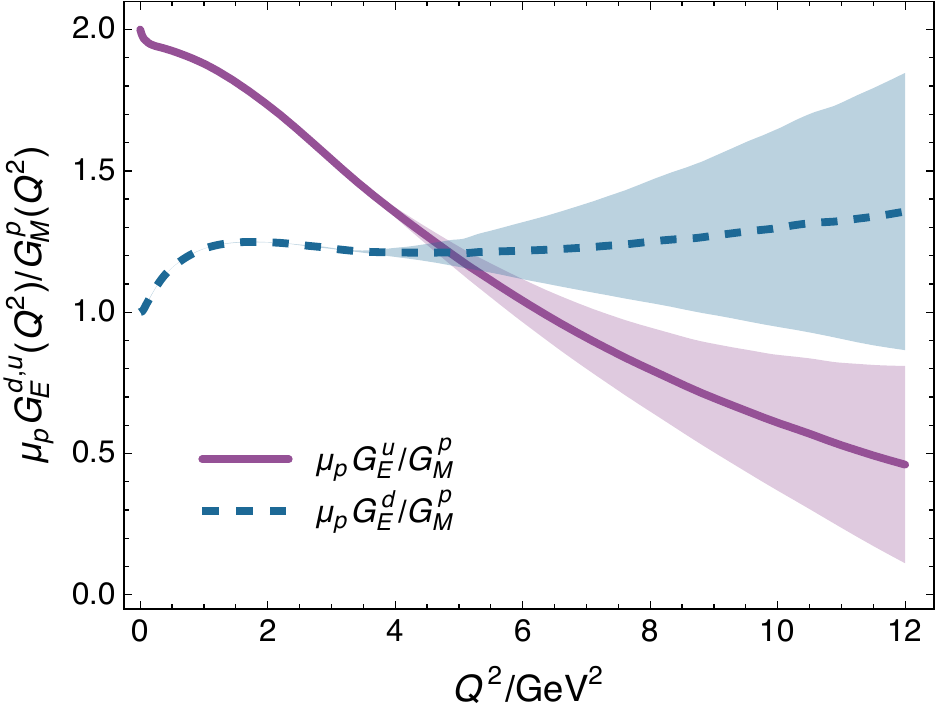}
\centering
\caption{\label{FigFlavSepGEM}
Flavour separation of the charge and magnetisation form factors, with each function normalised by $G_M^p$ in order to highlight their differing $Q^2$-dependence.}
%
\end{figure}

As explained elsewhere \cite{Segovia:2014aza}, in the isospin symmetry limit, the behaviours of $\mu_p G_E^p(Q^2)/G_M^p(Q^2)$ and $\mu_n G_E^n(Q^2)/G_M^n(Q^2)$ are not independent.  This is readily seen by exploiting isospin symmetry in writing a flavour separation of the charge and magnetisation form factors ($e_u=2/3$, $e_d=-1/3$):
\begin{equation}
\label{FlavSepGEM}
G_E^p  = e_u G_E^{pu} + e_d G_E^{pd}\,, \quad
G_E^n  = e_u G_E^{pd} + e_d G_E^{pu}\,.
\end{equation}
Regarding these identities, we refer to Fig.\,\ref{FigFlavSepGEM} and note that $G_E^p$ possesses a zero because, although remaining positive, $G_E^{pu}/G_M^p$ falls steadily with increasing $Q^2$ whereas $G_E^{pd}/G_M^p$ is positive and approximately constant.
On the other hand and consequently, $G_E^n$ does not exhibit a zero because $e_u>0$, $G_E^{pd}/G_M^p$ is large and positive, and $|e_d G_E^{pu}|$ is always less than $e_u G_E^{pd}$.

The character of $G_E^{pd}/G_M^p$ owes to the fact that $F_2^d$ is negative definite on the entire domain displayed in Fig.\,\ref{FlavSep}
-- because $F_2^n$ is negative thereupon, see Eq.\,\eqref{FlavourSep} --
and $G_E^d=F_1^d - (Q^2/[4 m_N]^2]) F_2^d$, whereas $F_1^d$ falls toward its zero from above..
This is not the case for the quark + diquark calculation, in which $F_2^d$ also exhibits a zero; so, at some $Q^2$, $G_E^{pd}$ begins to diminish in magnitude -- see, \emph{e.g}., Ref.\,\cite[Fig.\,7.3]{Cloet:2013jya}.
Plainly, the larger $Q^2$ behaviour of $F_2^d$ is key to the existence/absence of a zero in $G_E^n$.

Notwithstanding these differences, it is clear that, as in the quark + diquark picture \cite{Cui:2020rmu}, if the zero in $G_E^p/G_M^p$ moves to larger $Q^2$, then $G_E^n/G_M^n$ exhibits slower growth on $Q^2 \gtrsim Q^2_{F_1^d{\rm -zero}}$.
This correlation is also consistent with results obtained using lQCD \cite{Kallidonis:2018cas}.

Somewhat parenthetically, it is interesting to observe that, using the meson bound-state analogue of the approach employed herein \cite{Xu:2019ilh}, both the charged $\rho\,$- and $K^\ast$-meson electric form factors are predicted to exhibit a zero, whereas no zero is predicted in the neutral-$K^\ast$ form factor.
The explanation for the absence of a zero in the neutral-$K^\ast$ electric form factor \cite{Xu:2019ilh} is similar to that presented for $G_E^n$.
Notably, relocating the zero in $G_E^\rho$ by the ratio $m_p^2/m_\rho^2$, it is placed at $9.4(3)\,$GeV$^2$, within the domain defined by Eqs.\,\eqref{ZeroLocate}.
Furthermore, the electric form factor of the $J=1$ deuteron also exhibits a zero \cite{Kohl:2008zz}.
These remarks highlight that it is perhaps typical for the electric form factor of an electrically charged $J\neq 0$ bound state to possess a zero, owing to the potential for destructive interference between the leading charge form factor and magnetic and higher multipole form factors -- see, \emph{e.g}., Eq.\,\eqref{Sachs}.
This is not the case for $J=0$ \cite{Yao:2024drm} because such systems have only one electromagnetic form factor, $F_{J=0}$, and both valence contributions to $F_{J=0}$ are alike in sign.

\smallskip

\section{Summary and Perspectives}
\label{Epilogue}
Using a 
symmetry-preserving 
truncation of the quantum field equations relevant to calculation of hadron masses and interactions, this study delivers parameter-free predictions for all nucleon
charge and magnetisation distributions
and their flavour separation.
Each element in this analysis possesses an unambiguous link with analogous quantities in quantum chromodynamics (QCD) and the study unifies nucleon properties with those of numerous other hadrons -- see, \emph{e.g}., Refs.\,\cite{Eichmann:2020oqt, Ding:2022ows, Yao:2024drm}.
These features provide support for the reliability of the results herein.

The proton electric form factor, $G_E^p(Q^2)$, is predicted to possess a zero at a $Q^2$ location within reach of modern experiments [Fig.\,\ref{mupGEpGMp}A and Eq.\,\eqref{GEpzero}].  On the other hand, the neutron electric form factor, $G_E^n$, does not exhibit a zero [Fig.\,\ref{mupGEpGMp}B].  Consequently, anticipated experiments will see $|G_E^n/G_E^p|>1$, \emph{i.e}., an electric form factor of the charge-neutral neutron that is greater than that of the charge-one proton.  As revealed by a form factor flavour separation, these outcomes rest with the behaviour of the proton's $d$-quark Pauli form factor [Sec.\,\ref{NFFFS}].  Each of the highlighted form factor features are sensitive expressions of emergent phenomena in QCD.

No material improvement of the analysis herein can be anticipated before a way is found to include higher-order truncations in the continuum baryon bound-state problem or lattice-regularised QCD produces precise results on a similar domain to that discussed herein.  Meanwhile, the framework used herein can be applied to other high-profile challenges \cite{BESIII:2020nmeX, Anderle:2021wcyX, AbdulKhalek:2021gbhX, Quintans:2022utc, Carman:2023zke, Mokeev:2023zhq}, \emph{e.g.}, prediction of
baryon electroweak form factors,
nucleon-to-resonance transition form factors,
and nucleon gravitational form factors.
Such studies are underway.

\medskip

\noindent{\bf Declaration of competing interest}\,---\,%
The authors declare that they have no known competing financial interests or personal relationships that could have appeared to influence the work reported in this paper.

\medskip

\noindent{\bf Data availability}\,---\,%
This manuscript has no associated data or the data will not be deposited. [Authors’ comment: All information necessary to reproduce the results described herein is contained in the material presented above.]

\medskip

\noindent{\bf Acknowledgments}\,---\,%
We are grateful for constructive interactions with
P.\ Cheng,
L.\ Liu,
S.-X.\ Qin
and
Z.-N.\ Xu.
Work supported by:
National Natural Science Foundation of China (grant no.\ 12135007);
Natural Science Foundation of Jiangsu Province (grant no.\ BK20220122);
and STRONG-2020 ``The strong interaction at the frontier of knowledge: fundamental research and applications” which received funding from the European Union's Horizon 2020 research and innovation programme (grant agreement no.\ 824093).


\end{document}